%% file: main_JINST.tex
\documentclass[a4paper,11pt]{article}

\usepackage{jinstpub} 
\usepackage{multirow}
\usepackage{placeins}
\usepackage{titlesec}
\begin{document}

\title{\boldmath Beam Test of the First Prototype of SiPM-on-Tile Calorimeter Insert for the Electron-Ion Collider \\ Using 4 GeV Positrons at Jefferson Laboratory}
\author[a,b]{Miguel Arratia}
\author[a]{Bruce Bagby}
\author[a]{Peter Carney}
\author[a]{Jiajun Huang}
\author[a]{Ryan Milton}
\author[a]{Sebouh J. Paul}
\author[a]{Sean Preins} 
\author[a]{Miguel Rodriguez} 
\author[a]{Weibin Zhang}

\date{\today}
\affiliation[a]{Department of Physics and Astronomy, University of California, Riverside, CA 92521, USA}
\affiliation[b]{Thomas Jefferson National Accelerator Facility, Newport News, Virginia 23606, USA}

\emailAdd{miguel.arratia@ucr.edu}
\abstract{We recently proposed a high-granularity calorimeter insert for the Electron-Ion Collider (EIC) that uses plastic scintillator tiles read out by SiPMs. Among its innovative features are an ASIC-away-of-SiPM strategy for reducing cooling requirements and minimizing space use, along with employing 3D-printed frames to reduce optical crosstalk and dead areas. To evaluate these features, we built a 40-channel prototype and tested it using a 4 GeV positron beam at Jefferson Laboratory. The measured energy spectra and 3D shower shapes are well described by simulations, confirming the effectiveness of the design, construction techniques, and calibration strategy. This constitutes the first use of SiPM-on-tile technology in EIC detector designs.}

\maketitle

\input{Introduction}
\input{Results.tex}
\FloatBarrier

\input{Conclusions.tex}

 \section*{Acknowledgments}
  We thank members of the California EIC consortium and the ePIC Collaboration, and in particular Oleg Tsai, for valuable feedback related to our design and studies. We thank Alexander Somov for his guidance during our test at Hall D and for his continued support afterward. We thank Ron Soltz and Ernst Sichtermann for supporting and guiding Jiajun Huang, and Miguel Rodriguez. 
  
  This work was supported by MRPI program of the University of California Office of the President, award number 00010100. Sebouh J.~Paul acknowledges support by the Jefferson Laboratory EIC Center Fellowship. Sean Preins was supported by a HEPCAT fellowship from DOE award DE-SC0022313. This work is based upon work supported by the U.S.~Department of Energy, Office of Science, Office of Nuclear Physics, RENEW under Award Number DE-SC0022526, which supported Peter Carney, Jiajun Huang, and Miguel Rodriguez. Miguel Arratia acknowledges support through DOE Contract No. DE-AC05-06OR23177 under which Jefferson Science Associates, LLC operates the Thomas Jefferson National Accelerator Facility.

  \newpage
\bibliographystyle{utphys} 
\bibliography{biblio.bib}

\end{document}

%% file: Introduction.tex
\section{Introduction}
The future Electron-Ion Collider (EIC)~\cite{Accardi:2012qut} aims to explore nuclear structure and dynamics across a broad range of kinematics. In order to achieve this goal, a large acceptance detector called ePIC is being developed, following the designs outlined in Refs.~\cite{ECCE,ATHENA:2022hxb}. Maximizing acceptance in the ePIC central detector, which nominally spans the range $-4.0< \eta <4.0$ to fulfill the EIC Yellow Report requirement~\cite{AbdulKhalek:2021gbh}, poses challenges due to the EIC's 25 mrad beam-crossing angle. This angle leads to a complex beampipe geometry, particularly in the vicinity of the ePIC forward calorimeter.

This challenge motivated the development of the high-granularity calorimeter insert (CALI)~\cite{hcalInsert}, especially designed to cover the range $3.2 < \eta < 4.0$. The design incorporates absorber layers with unique shapes to accommodate the complex beampipe. Additionally, this design offers high granularity to enhance performance in measuring jets, and to manage radiation damage and beam-gas interactions effectively.

The CALI is based on SiPM-on-tile technology~\cite{Blazey:2009zz,Simon:2010hf}, offering flexibility, scalability, cost-effectiveness, and high performance for high-granularity calorimeters~\cite{RevModPhys.88.015003}. In recent years, the SiPM-on-tile approach has become a popular option for various experiments~\cite{CALICE:2010fpb,CALICE:2022uwn,Dannheim:2019rcr,CEPCStudyGroup:2018ghi,Li:2021gla,Jiang:2020tve,Duan:2021mvk,ILDConceptGroup:2020sfq,CMS:2017jpq}. A significant application of this technology is in the HGCAL upgrade for CMS at the high-luminosity LHC~\cite{Contardo:2020886,CMS:2017jpq}. Numerous beam-test studies conducted by the CALICE Collaboration~\cite{RevModPhys.88.015003}, as well as more recent studies by the CMS Collaboration~\cite{Belloni:2021kcw,CMS:2022jvd}, have provided invaluable references for new designs. 

Unlike the designs in CMS HGCAL~\cite{CMS:2017jpq} and CALICE AHCAL~\cite{CALICE:2022uwn}, the CALI will have its ASIC readout chips at the back of the device, up to a meter away from the SiPMs, instead of directly on them. This configuration arises from cooling constraints and limitations on longitudinal space within the ePIC detector. These chips, likely modified versions of the CMS HGROC ASIC~\cite{Bombardi2020HGCROCSiAH}, will get the SiPM pulses via a PCB, as initially proposed in Ref.~\cite{hcalInsert}. The PCB will serve as a shielded cable that runs longitudinally to the back of the calorimeter. This analog signal transport method for SiPM pulses could potentially degrade its performance. One aim of the work presented in this paper is to investigate this aspect of the CALI design.

The CALI will cover an area of 60$\times$60~cm$^{2}$ and consist of 64 iron-scintillator layers. Each cell is placed in 3D-printed frames that define a layer, simultaneously holding the scintillator cells in place and reducing optical crosstalk between them~\cite{Arratia:2023rdo}. Unlike the CMS and CALICE designs~\cite{CMS:2017jpq,CALICE:2022uwn}, the entire layers of the CALI will be sandwiched between a pair of Enhanced Specular Reflector (ESR) foils, and the edges of the cells will be coated with reflective paint. Each layer will be positioned between a plastic cover and a PCB with SiPMs. This 3D-printed frame approach was introduced and tested on the bench in Ref.~\cite{Arratia:2023rdo}, and it was tested under beam conditions for the first time, as reported here.

In this paper, we present the outcomes of building the first prototype of the CALI design and conducting a test-beam experiment with a positron beam at the Thomas Jefferson National Accelerator Facility (JLab). The objectives of this test were to validate the CALI design, particularly its incorporation of 3D-printed frames and the ASIC-away-of-SiPM  approach, verify its simulations, assess performance characteristics, refine construction methods, and acquire experience in operating and calibrating a SiPM-on-tile detector.

This paper is structured as follows: Section~\ref{sec:prototype} describes the prototype, Section~\ref{experimentalsetup} outlines the test-beam setup, Section~\ref{sec:dataanalysis} details the data analysis, Section~\ref{sec:results} presents the results, and Section~\ref{sec:summary} offers a summary.

\section{Prototype}
\label{sec:prototype}
\begin{figure}[h!]
    \centering
    \includegraphics[width=0.6\linewidth]{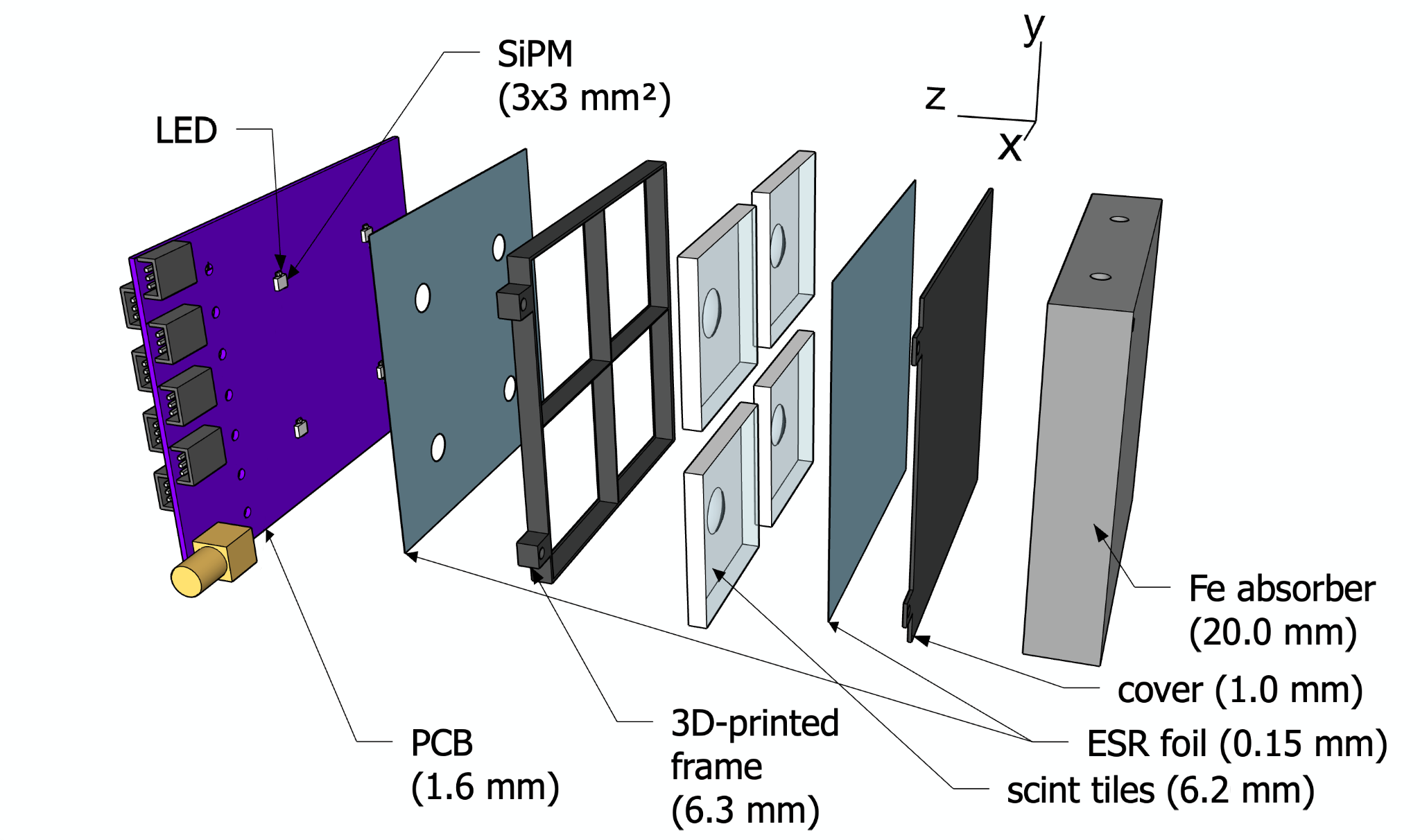}
    \includegraphics[width=0.39\textwidth]{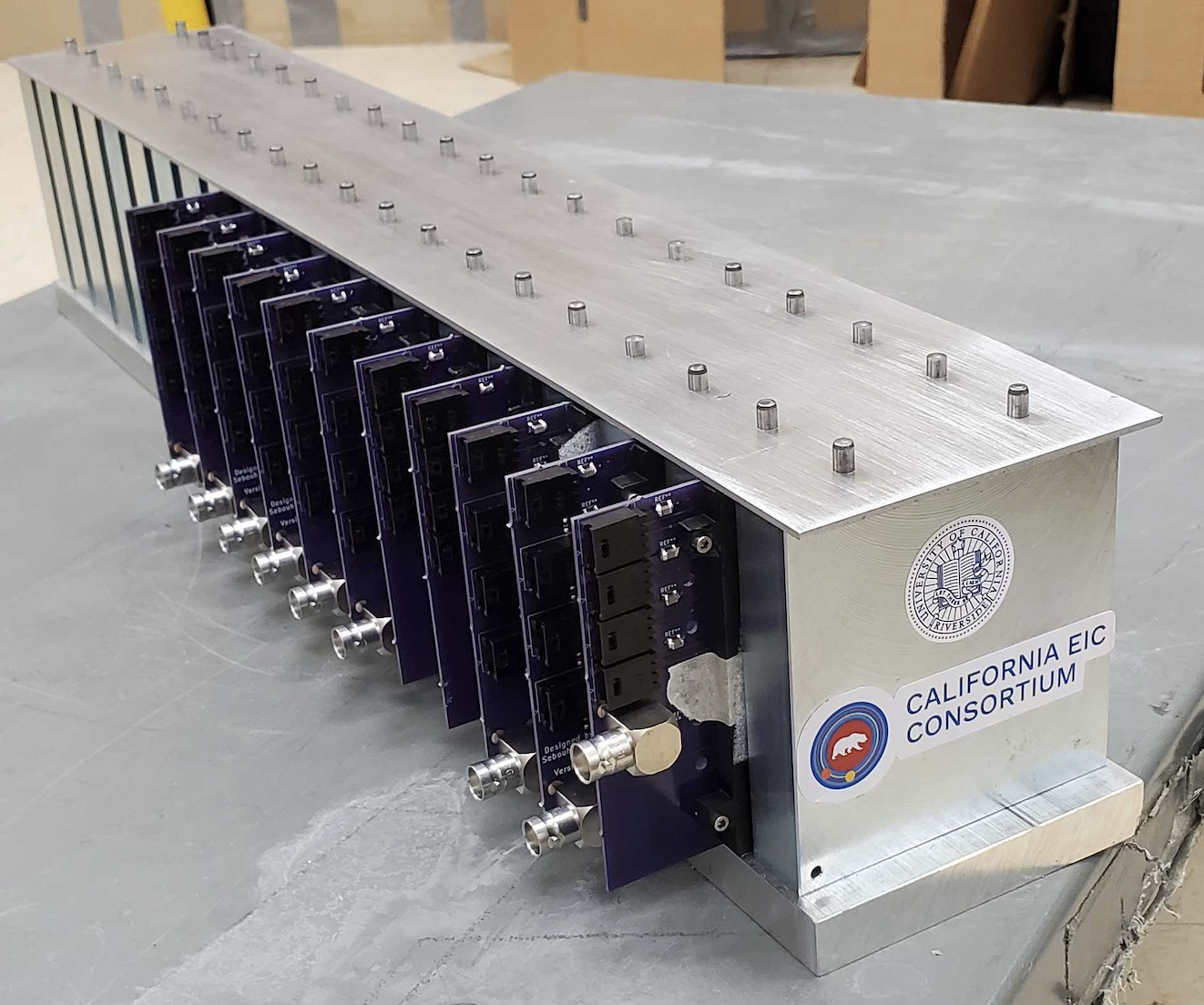}
    \caption{Left: exploded view of prototype layer design. Right: assembled Calorimeter Insert prototype.}
    \label{fig:Prototype}
\end{figure}

The first CALI prototype consisted of 10 sampling layers, illustrated in Figure~\ref{fig:Prototype}, with a transverse active area of 9.5$\times$ 9.5~cm$^2$. Each steel block was 2~cm thick, corresponding to 1.1 $X_{0}$, resulting in a total radiation length of 11.7 $X_{0}$. The tiles were made of Bicron 404 plastic scintillator, each with a thickness of 6.2 mm~\footnote{The CALI will use 3 mm thick scintillator, but we opted for this thickness because we had the material on hand.} and a dimple in its center. The dimples were crafted using a CNC machine and subsequently hand-polished with sandpaper and NOVUS polishing liquid.

\FloatBarrier
Following our earlier R\&D~\cite{Arratia:2023rdo}, we coated the edges of each cell with white reflective paint (Saint Gobain BC-621) and fit them into a plastic frame produced by an FDM 3D printer (Ender 3 V2 with a $0.4$~mm nozzle size). The plastic frame had a width of 0.4 mm separating each scintillating tile from its neighbors to reduce optical crosstalk. The layers are sandwiched between $0.15$~mm ESR foils, and the layer of ESR foil that came into contact with the PCB had circular holes cut into it to couple to the SiPM. We employed a 1.6 mm PCB to mount the SiPMs and form the foundation of each layer, as illustrated in Fig. \ref{fig:Prototype}. Each PCB had four SiPMs soldered onto it. We used the HPK S14160-3015PS SiPM, which has a 3$\times$3 mm$^{2}$ area and contains 39,984 pixels~\cite{SiPM}.

Among the 10 sampling layers, the first four layers (0-3) included four square scintillator tiles, each measuring 22 cm$^{2}$. The last six layers (4-9) featured four hexagonal tiles, each measuring 13 cm$^{2}$. We tested both cell shapes for comparison purposes. Although the hexagonal tiles do not tessellate the area efficiently in this prototype, the CALI design has the potential to incorporate hexagonal tessellations with minimal or no dead area by using smaller cells~\cite{hcalInsert}, and a staggered design~\cite{Paul:2023okc}. The centers of the dimples in the cells were aligned with the SiPMs on the boards within 1 mm, constrained by the dimensions of the cells and the frames that housed them.

With the scintillating tiles in place within the frames, the frames were then fastened to a non-active region of the PCB, extending from the left side of the prototype as shown in Fig.\ref{fig:Prototype}. Each of the sampling layers was then situated between the steel absorbers, also depicted in Fig.\ref{fig:Prototype}. These absorbers were held in place by top and bottom plates linked with dowel pins. Lastly, the prototype was housed in a dark box made out of cardboard. 

\section{Experimental Setup and Data Acquisition System}
\label{experimentalsetup}
The beam test was conducted in January 2023 at Hall D in JLab, using positrons from the Pair Spectrometer (PS) of the GlueX experiment~\cite{GlueX:2020idb}. The data-acquisition and trigger system we used was the CAEN FERS-5200 unit (model DT5202), which was entirely independent from the GlueX PS system. This configuration prevented us from using the PS hodoscope~\cite{Barbosa:2015bga} for triggering and tagging the positron energy event-by-event. Instead, we used the prototype's location and size to estimate exposure to a beam with an energy range of about 4$\pm$1 GeV. 

The FERS-5200 unit, which contains two CITIROC 1A ASICs, can bias and read out up to 64 SiPMs independently~\cite{Venaruzzo:2020usj}, and it also provides self-triggering capabilities. The CALI prototype PCBs are connected to the FERS-5200 unit through 1-meter shielded cables using AMPMODU type 102241-1 connectors (CAEN A5261~\cite{A5261}). This setup roughly mimics the ASIC-away-of-SiPM configuration that is to be used in the CALI design~\cite{hcalInsert}.

The FERS-5200 unit was set to operate in ``Spectroscopy mode,'' which, upon triggering, performs simultaneous acquisition across all channels using a 13-bit analog-to-digital conversion and allows for a maximum trigger rate of 100 kHz. The trigger logic mandated that a minimum of four channels exceed a threshold of 220 ADC units\footnote{The trigger logic was configured with ``MajorityLevel'' set to 4, ``TD\_CoarseThreshold'' at 220, and ``QD\_CoarseThreshold'' at 250; the gain settings for both low and high gain were also set to 50.}. The SiPMs were biased at 43 V, which corresponds to about 5 V over-voltage.  The readout data were synchronized to a laptop for storage and analysis. The max positron rate observed was about 3 kHz. We collected \(6.2 \times 10^5\) trigger events.

\section{Data Analysis}
\label{sec:dataanalysis}

\subsection*{Pedestal and Cosmic-Ray Runs}
Dedicated pedestal runs, obtained with random triggers in beam-off conditions, were conducted to determine the pedestal mean position and widths for each channel independently. The ADC count spectrum in each channel was fit to a Gaussian distribution. The average values of the pedestal means and widths were 59.6 ADC counts and 3.2 ADC counts, respectively.

Cosmic rays were used to calibrate the prototype on a channel-by-channel basis. Before the complete assembly in Hall D, the sampling layers were arranged in a vertical stack to maximize their exposure to cosmic rays. The layers were situated on top of each other in a modular 3D-printed rack. We conducted the cosmic ray collection overnight, using the same FERS-5200 settings as in the beam test. The energy spectra, measured in ADC counts, were well-described by a Landau distribution. After subtracting the pedestal, the most probable value from a Landau fit was used as the MIP scale for each channel individually.

The MIP scale varied from channel to channel, falling within the range of approximately 55--70 ADC counts for the first four layers and 80--120 ADC counts for the subsequent six layers. These ranges are associated with the larger square tiles and the smaller hexagonal tiles, respectively. The primary factor contributing to this discrepancy is likely the difference in area between the two tile shapes; a larger area results in a longer average path for a photon to reach the SiPM, leading to greater signal attenuation and, consequently, a smaller light yield. The relatively wide fluctuation of MIP scale values within the same cell geometry can be attributed to inconsistencies in the polishing process.

\subsection*{Energy Calibration}
\label{sec:energycalibration}
The measured energy was calibrated from the ADC scale to the MIP scale as follows:
\begin{equation}
    E_{i} [\mathrm{MIP}]  =  \frac{\left ( E_{i} [\mathrm{ADC}] - \langle\mathrm{pedestal}\rangle_{i} [\mathrm{ADC}] \right )}{ \mathrm{MPV}_{i} [\mathrm{ADC/MIP}]}
\end{equation}
Here \( E_{i}[\mathrm{ADC}] \) is the measured charge for channel $i$ in ADC scale before pedestal subtraction, \( \langle \mathrm{pedestal}\rangle_{i} [\mathrm{ADC}] \) is the pedestal mean for channel $i$, and \( \mathrm{MPV} _{i}[\mathrm{ADC/MIP}] \) is the $i$-th channel most-probable value energy deposition in ADC units by a MIP after pedestal subtraction.

\subsection*{Simulation}
We used the \textsc{DD4HEP} framework \cite{DD4HEP} to simulate the response of the CALI prototype to positrons and muons. The version of \textsc{Geant4} \cite{GEANT4} used within \textsc{DD4HEP} is 11.1.1, along with the \textsc{FTFP\_BERT} physics list and Birks' constant set to 0.126~mm/MeV.

The simulated hits were digitized using a 13-bit ADC, incorporating the average and RMS pedestal values extracted from data, but without including any electronic or optical crosstalk. The MIP scale was determined with the simulation of muons in a manner analogous to how it was determined in the data. This calibration was assumed to be identical for all channels. 

\begin{figure}[h!]
    \centering
    \includegraphics[width=0.75\textwidth]{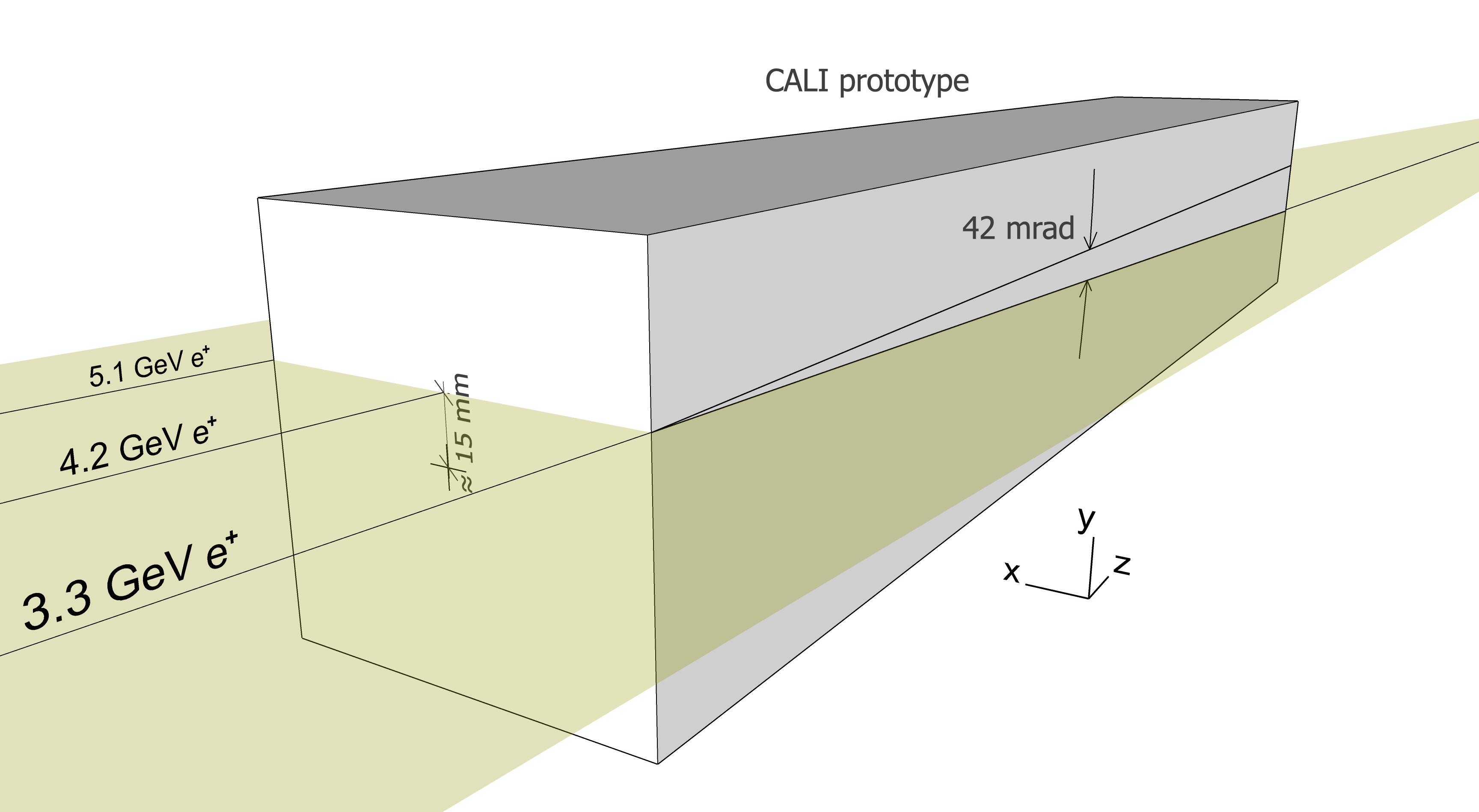}
    \caption{Relative position and tilt of the beam midplane in relation to the prototype, as identified by adjusting the simulation to align with the data.  Also indicated are the energies of positrons that hit the prototype at the left edge, middle, and right edges of its front face.}
    \label{fig:tilt}
\end{figure}

We tuned the simulation parameters to match the beam conditions. Since our setup lacked a tracking system, we couldn't accurately determine the beam's direction. Analysis of the prototype data revealed a slope in the energy-weighted vertical shower position as a function of layer number, suggesting a beam tilt and a vertical shift relative to the prototype's center. To address this, we adjusted the shift distance and tilt angle to match the data. Our findings indicated a vertical misalignment and tilt polar angle of approximately 15 mm and $40 < \theta < 44$ mrad, respectively, as illustrated in Fig. \ref{fig:tilt}. The beam energy was parameterized to exhibit linear dependence on the azimuthal angle, as expected from the the pair spectrometer that generated the positrons. We set the beam energy to range from 3.3 GeV to 5.1 GeV across the prototype's front face, roughly aligning with expectations based on its location and transverse size.

\subsection*{Hit and Event Selection}
The trigger logic used during data collection was also imposed on our simulated data set. For the following analysis, we only consider hits with $E > 0.3$ MIP, and select events with at least 4 hits above this threshold. The same hit and event selection criteria were applied to the simulation. This cut removed 0.2\% of events from the data and 0.9\% of the events from the simulation.

%% file: Results.tex
\section{Results}
\label{sec:results}
\begin{figure}[h!]
    \centering
    \includegraphics[width=.90\textwidth]{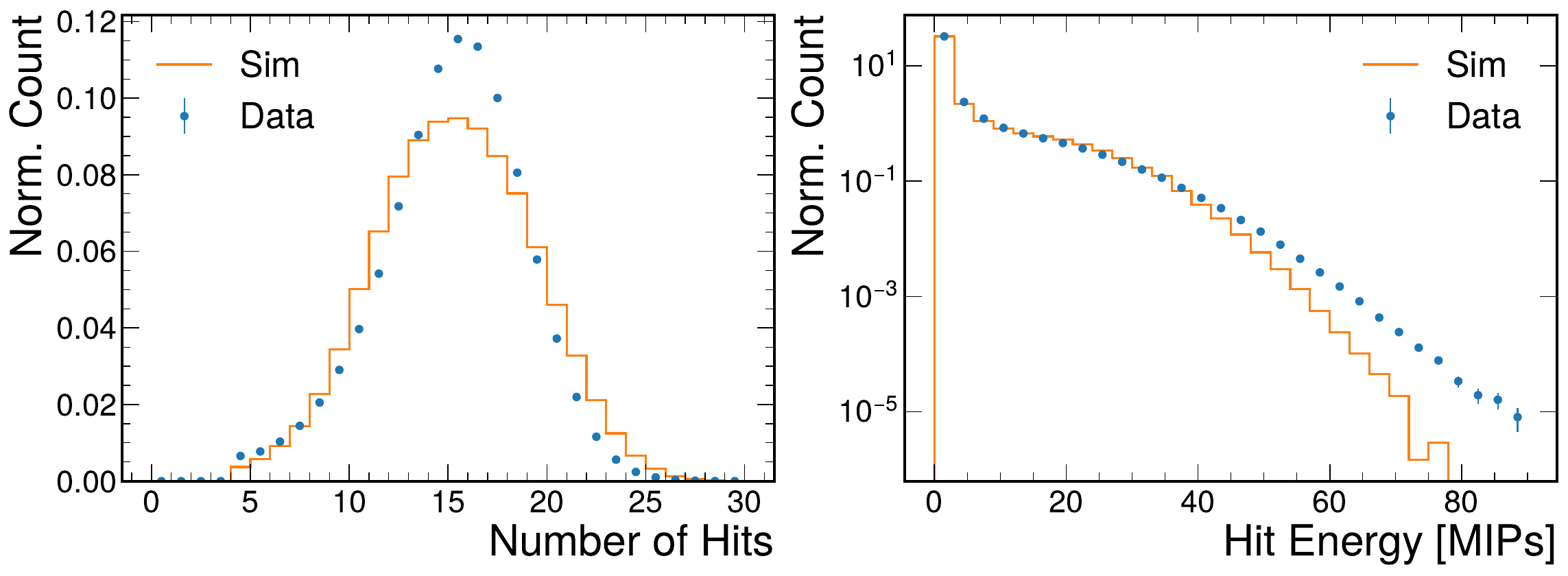}
    \caption{Number of hits above threshold per shower (left) and corresponding hit-energy spectrum (right).}
    \label{fig:hit}
\end{figure}

Figure~\ref{fig:hit} presents the distribution of the number of hits above the $E > 0.3$ MIP threshold, along with the hit-energy spectra. Most events show a hit multiplicity ranging between 10 and 20, with a peak at approximately 15 hits per shower in both the data and simulation. However, the simulation exhibits a broader distribution. The hit-energy spectra show that the energy of a single hit can reach up to 80 MIPs. A good agreement between the data and simulation is observed in the low-energy and mid-energy region but not in the high-energy region.

\begin{figure}[h!]
    \centering
    \includegraphics[width=1.0\textwidth]{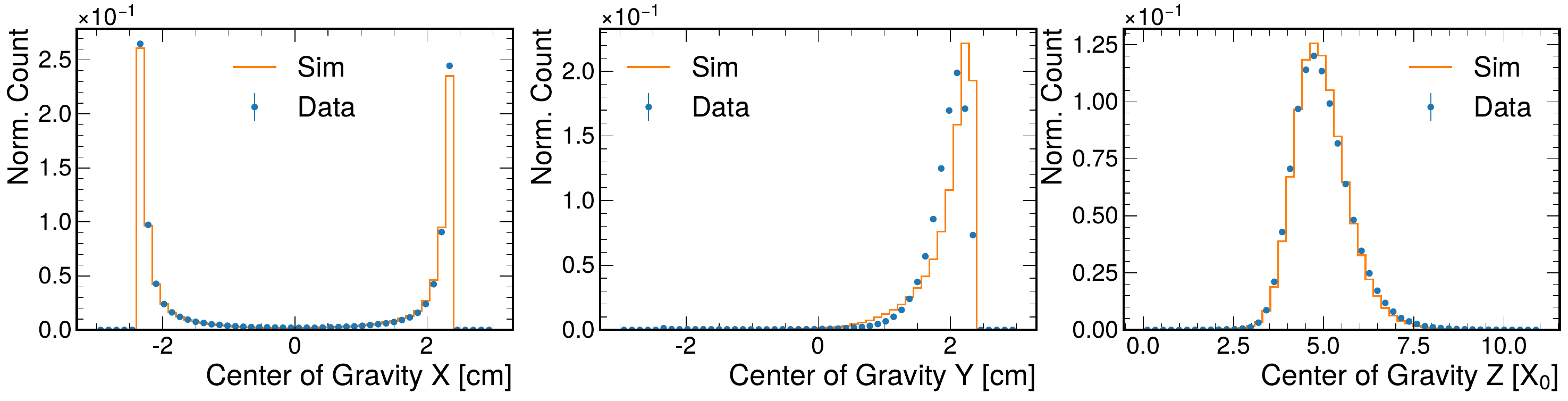} 
    \caption{Shower center of gravity in horizontal (left), vertical (middle), and longitudinal (right) directions.}
    \label{fig:COG}
\end{figure}

The tilted and shifted beam condition is particularly noticeable in the Center of Gravity (COG) observable, which is defined as a 3D vector and calculated event-by-event as follows:
\begin{equation}
    \text{COG} = \frac{\sum_{i} E_{i} \cdot \vec{X}_{i}}{\sum_i E_{i}} 
\end{equation}
Here, the index $i$ cycles through all hits with $E > 0.3$~MIP. $E$ denotes the hit energy, while $\vec{X}_{i}$ is the 3D position of the center of the tile where the $i$-th hit takes place. The horizontal, vertical, and longitudinal projections of the COG are presented in Figure~\ref{fig:COG}. The transverse variables are expressed in units of cm, and the longitudinal variables are in units of radiation lengths.

The $\text{COG}_x$ distribution shows a nearly symmetrical pattern, a feature accurately captured by the simulation. In contrast, the $\text{COG}_y$ distribution shows a single peak that suggests an upward shift in the beam's position, while the tail indicates a slight beam tilt. This distribution is not as well captured by the simulation, pointing to imperfections in modeling the beam directions. Meanwhile, the $\text{COG}_z$ distribution has a Gaussian-like shape, peaking at around 5 $X_0$ or approximately in the fourth layer, and is well described by the simulation.

\begin{figure}[h!]
    \centering
    \includegraphics[width=0.80\textwidth]{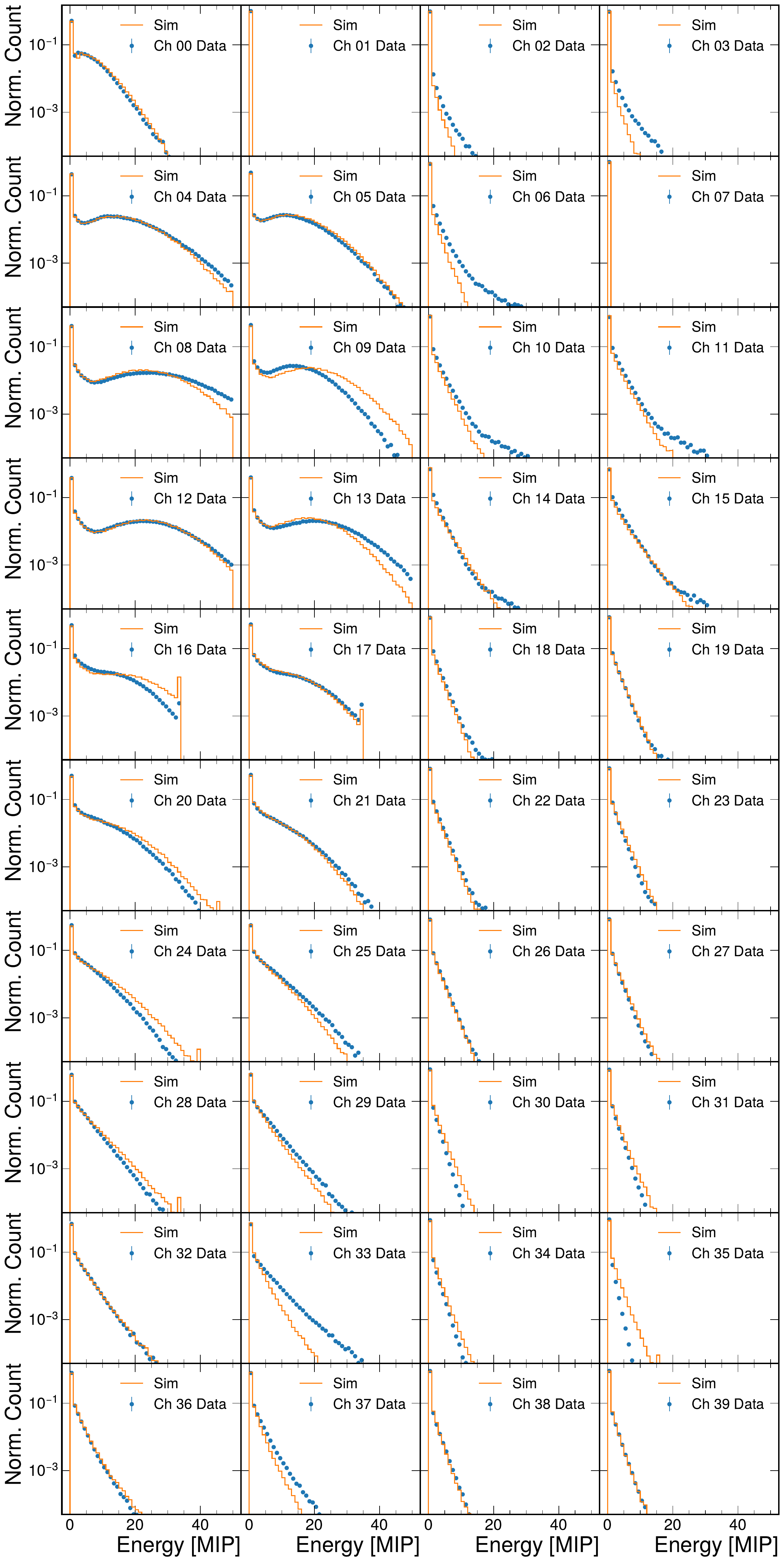}
    \caption{Energy spectra for each channel. Each column represents a single layer in the detector.}
    \label{fig:cell_comparison}
\end{figure}

Figure \ref{fig:cell_comparison} displays the hit-energy spectra across all 40 channels. Most spectra are reasonably described by the simulation. Channels 1 and 7 were non-functional, likely due to lost cable connections during transport to the beam test area, and are also masked in the simulation. The prominent spikes in channels 16 and 17 were due to a cap set by the CAEN unit's ADC dynamic range; this effect was included in the simulated data as well.

\FloatBarrier

Channels 9 and 16 show larger signals in the simulation than what is observed in the data. Considering that their horizontally symmetric counterparts, channels 8 and 17, align well with the simulation, a likely explanation for this discrepancy could be an incorrect MIP scale for these specific channels in the data. This variation in MIP scale could have resulted from the reassembly of the prototype between cosmic and beam runs, potentially shifting the tile positions relative to the SiPMs and affecting the light yields~\cite{deSilva:2020mak}. To address this, future tests will incorporate in-situ cosmic runs to overcome this specific challenge.

\begin{figure}[h!]
    \centering
    \includegraphics[width=0.99\textwidth]{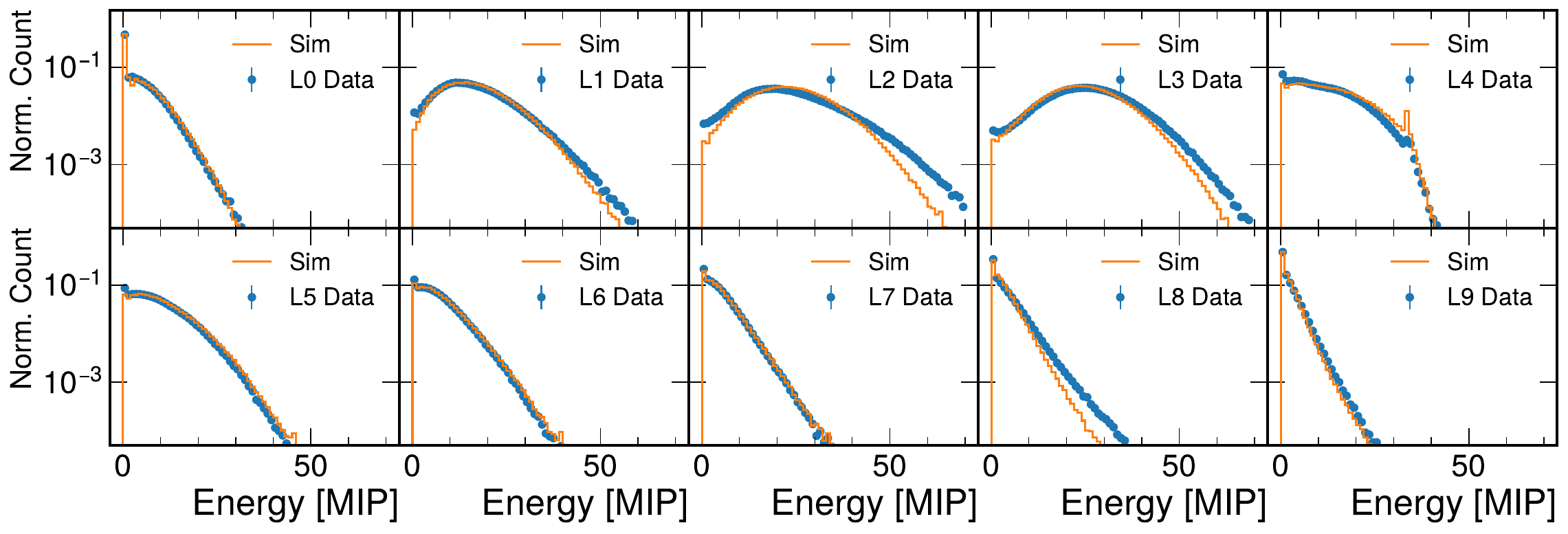}
    \caption{Total energy per layer.}
    \label{fig:layer_comparison}
\end{figure}

Given the reasonable agreement observed at the channel level, a similar level of congruence is anticipated at the layer level, as depicted in Fig.~\ref{fig:layer_comparison}. The most pronounced signals consistently occur in the first four layers. The kink seen in the layer 4 distribution can be attributed to the data capping effect observed in channels 16 and 17.

\begin{figure}[h!]
    \centering
    \includegraphics[width=0.52\textwidth]{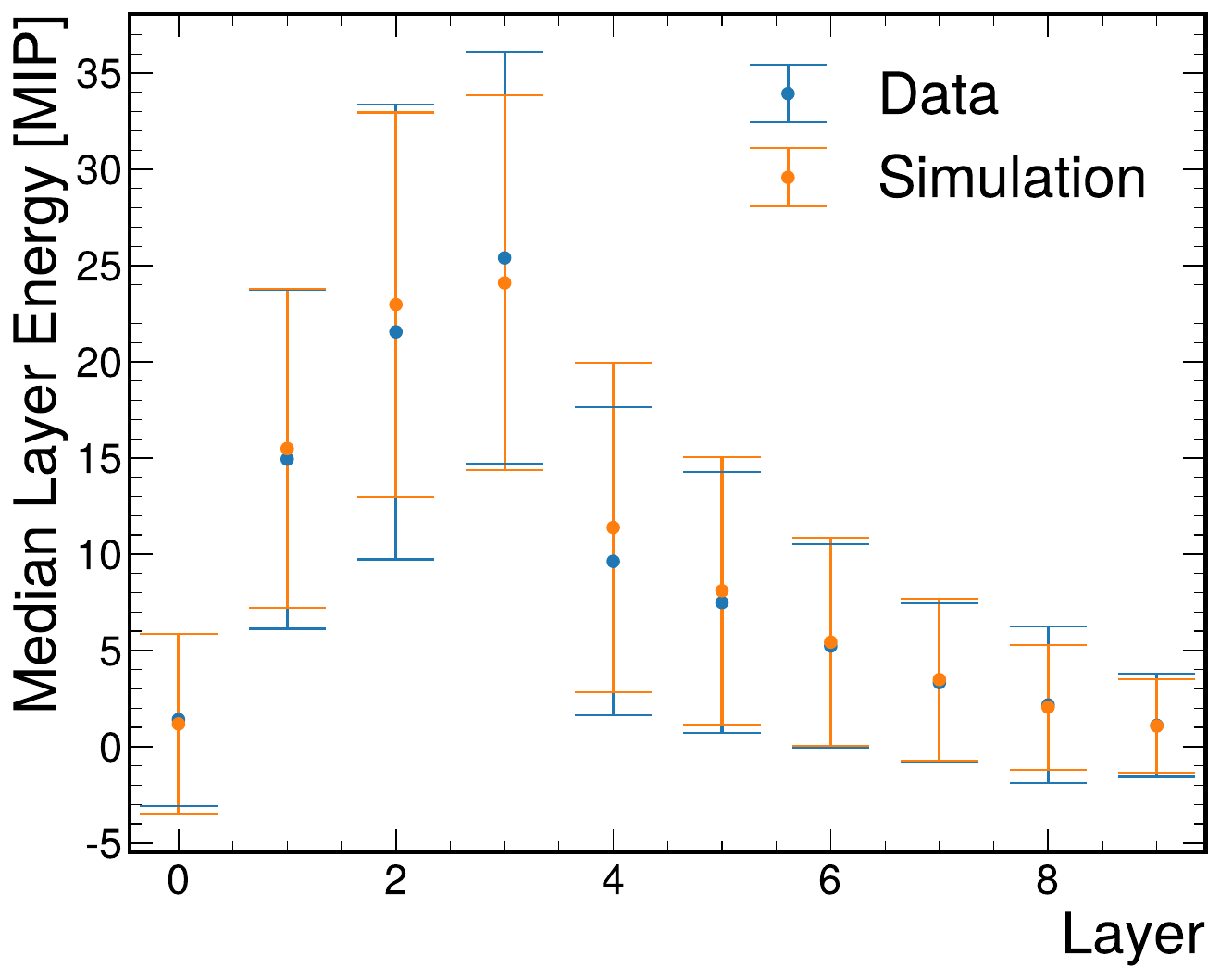}
    \caption{The median (markers) and RMS (error bar) of the energy distribution per each layer.}
    \label{fig:layer_median_energy_comparison}
\end{figure}

Figure~\ref{fig:layer_median_energy_comparison} displays the median values and RMS of the energy distribution for each layer. The measured median values align closely with the simulation, although the simulation slightly overestimates the values compared to those observed in the data. The RMS of the layer energy is also reasonably described, although the simulation results are somewhat smaller than those in the data.
\FloatBarrier

\begin{figure}[h!]
    \centering
    \includegraphics[width=0.45\textwidth]{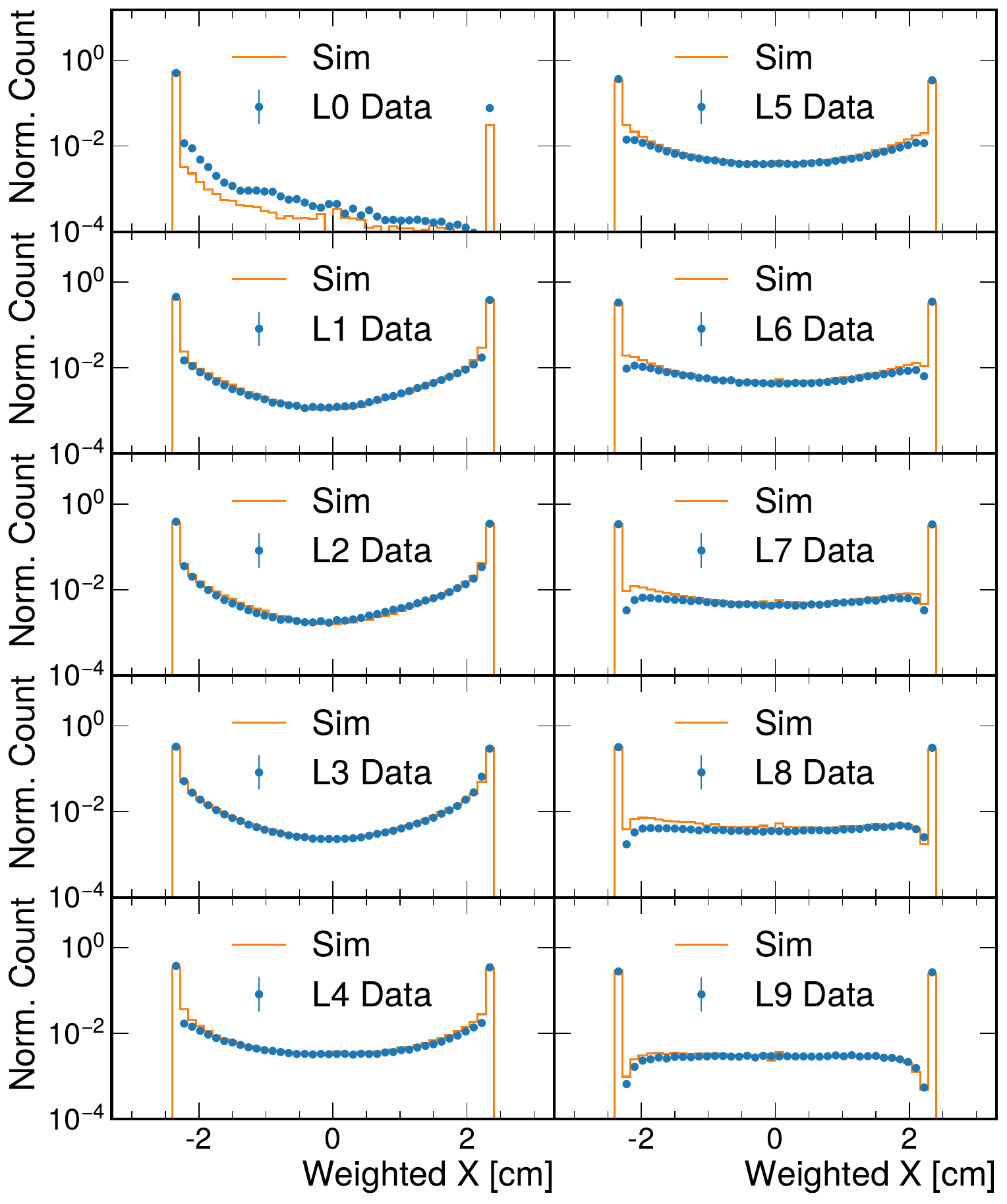} 
    \includegraphics[width=0.45\textwidth]{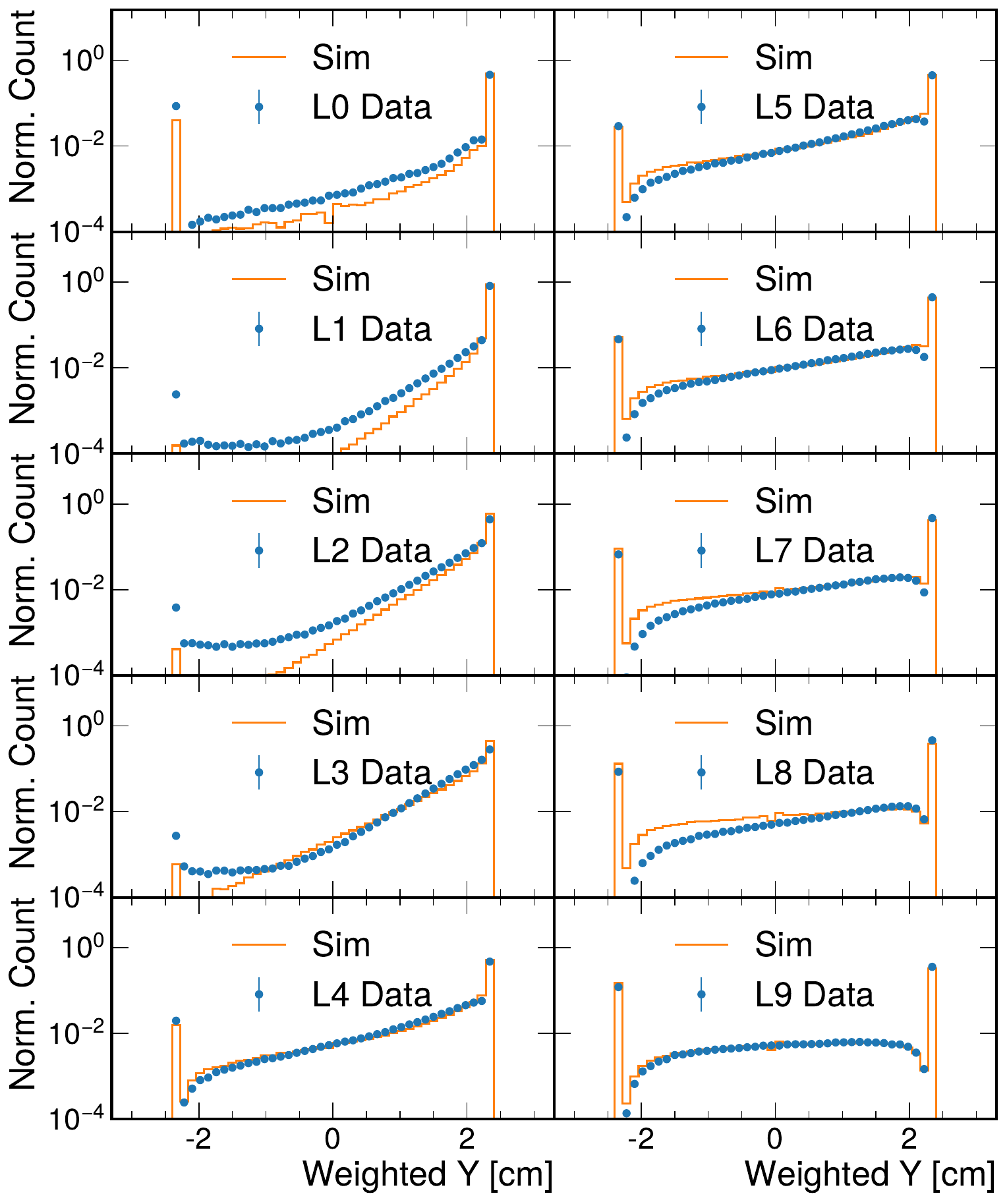}
    \caption{Energy-weighted $x$ (left two columns) and $y$ positions (right two columns) in each layer.}
    \label{fig:layer_xy_comparison}
\end{figure}

Figure~\ref{fig:layer_xy_comparison} shows that the energy-weighted hit positions for each layer align with the simulation reasonably well. The $x$-distributions suggest symmetric beam coverage along the $x$-axis, spanning the entire prototype. Conversely, the $y$-distributions reflect the beam experienced a shift and tilt in the $y$-direction. The central region between the two peak bins tends to flatten out as the sampling layers progress, as expected. Our simulation does not account for optical crosstalk, so the data-simulation agreement supports the notion that it is negligible in the data, consistent with Ref.~\cite{Arratia:2023rdo}. 

\begin{figure}[h!]
    \centering
    \includegraphics[width=0.48\textwidth]{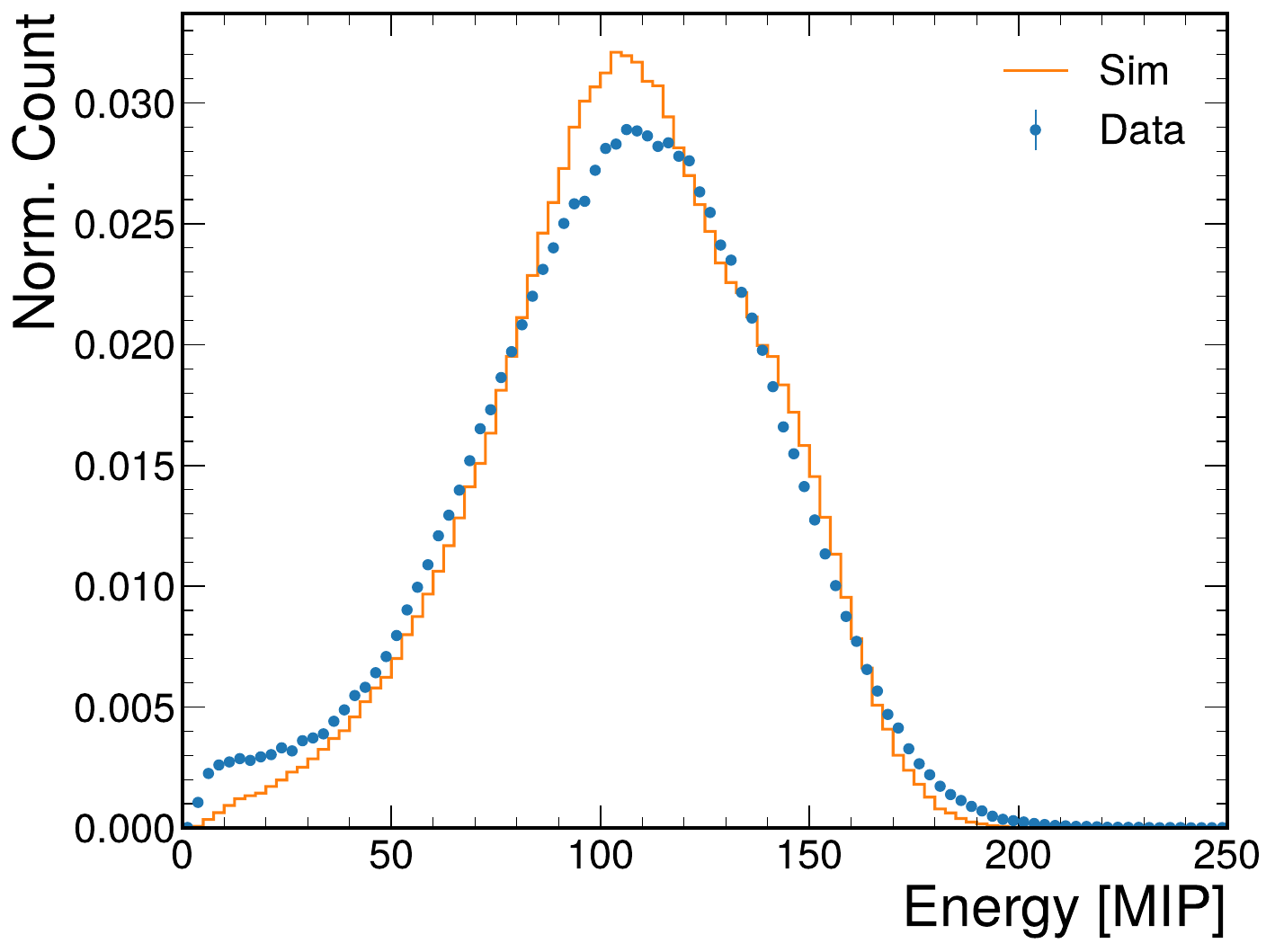}
    \caption{The total energy per shower.}
    \label{fig:event_comparison}
\end{figure}

Finally, the total energy spectrum of the shower is shown in Fig.~\ref{fig:event_comparison}. The simulation offers a reasonable approximation of the data, accurately capturing both its mean and standard deviation, which its driven by the beam-energy spread. These discrepancies may arise from inherent challenges in precisely emulating beam conditions, potential miscalibration, or cell non-uniformity. 

%% file: Conclusions.tex
\section{Summary and Outlook}
\label{sec:summary}
We conducted comprehensive studies on the first prototype for the EIC Calorimeter Insert~\cite{hcalInsert}, which is based on the SiPM-on-tile technology. A proof-of-concept 40-channel prototype was built and tested using positron beams at Jefferson Laboratory. Key metrics such as energy spectra and 3D shower shapes were assessed, verified against simulations, and found to be in reasonable agreement.

The test results suggest the potential feasibility of the ASIC-away-of-SiPM/SiPM-on-tile strategy in the CALI design, providing the potential to reduce cooling needs and make more efficient use of space. Additionally, shower shape analysis suggests negligible optical crosstalk between adjacent cells, validating the novel use of 3D-printed frames to define ``megatiles''. Our analysis also achieved the objective of demonstrating the prototype's effectiveness in detecting particle shower axes through 3D position and energy measurements.

These results were obtained with just a preliminary prototype featuring $O(1)$\% of the channel count planned for the final detector. Nonetheless, our results aligns with our objectives, confirming key aspects of the CALI design. This work has resulted in improved construction methods and more clearly defined strategies for both the operation and calibration of the CALI.

Importantly, we have established a precedent by demonstrating the pioneering use of SiPM-on-tile technology for EIC detectors. This offers insights that could inform future studies of other EIC subdetectors, such as the forward hadronic calorimeter~\cite{Bock:2022lwp}, the zero-degree calorimeter, and the few-degree calorimeter~\cite{Arratia:2023gyx}, and may even extend to experiments beyond the EIC.